\documentclass[aps,prl,reprint,groupedaddress,showpacs,floatfix]{revtex4-1}

\usepackage{graphicx}% Include figure files
\usepackage{dcolumn}% Align table columns on decimal point
\usepackage{bm}% bold math
\usepackage{amsmath}% bold math
\usepackage{amssymb}
\usepackage{natbib}

\begin{document}

\title
%  {Structural Phase Transition in Monolayer As$_{\sf 1-x}$P$_{\sf x}$ Compounds:
%  A Computational Study}
  {Structural Transition in Layered
  As$_{\sf\textbf{1-x}}$P$_{\sf\textbf{x}}$ Compounds:
  A Computational Study}

\author{Zhen Zhu}
\affiliation{Physics and Astronomy Department,
             Michigan State University,
             East Lansing, Michigan 48824, USA}

\author{Jie Guan}
\affiliation{Physics and Astronomy Department,
             Michigan State University,
             East Lansing, Michigan 48824, USA}

\author{David Tom\'{a}nek}
\affiliation{Physics and Astronomy Department,
             Michigan State University,
             East Lansing, Michigan 48824, USA}
\email{tomanek@pa.msu.edu}

%%%%%%%%%%%%%%%%%%%%%%%%%%%%%%%%%%%%%%%%%%%%%%%%%%%%%%%%%%%%%%%%%%%%%
%% The manuscript does not need to include \maketitle, which is
%% executed automatically.  The document should begin with an
%% abstract, if appropriate.  If one is given and should not be, the
%% contents will be gobbled.
%%%%%%%%%%%%%%%%%%%%%%%%%%%%%%%%%%%%%%%%%%%%%%%%%%%%%%%%%%%%%%%%%%%%%

\begin{abstract}
As a way to further improve the electronic properties of group V
layered semiconductors, we propose to form in-layer 2D
heterostructures of black phosphorus and grey arsenic. We use
\textit{ab initio} density functional theory to optimize the
geometry, determine the electronic structure, and identify the
most stable allotropes as a function of composition. Since pure
black phosphorus and pure grey arsenic monolayers differ in their
equilibrium structure, we predict a structural transition and a
change in frontier states, including a change from a direct-gap to
an indirect-gap semiconductor, with changing composition.

% We present the relative stability and electronic properties of the
% alloy system As$_{1-x}$P$_x$. Based on our calculations, we find
% by varying the ratio of phosphorus component $x$, structural phase
% transition could be induced between A17 type structure (black
% phosphorus) and A7 type structure (grey arsenic) near $x$=0.93. We
% also expect phase coexistence of both A17 and A7 structures within
% larger range of $x$ due to their negligible stability difference.
% In A17 phase, although the electronic band gap value almost does
% not vary, alloying between phosphorus and arsenic could change it
% from direct in black phosphorus) to indirect in black arsenic.
% Moreover, the frontier orbital of A7 phase will switch from lone
% pair electrons in $\beta$-P to the inlayer $\sigma$-bond electrons
% in grey arsenic.
\end{abstract}


\maketitle

\newpage
\section{Introduction}

Few-layer structures of group V elements, including phosphorene
and arsenene, are emerging as promising candidates for
two-dimensional (2D) electronic materials
application\cite{Narita1983,Maruyama1981,DT229,DT244,BiluLiu15}.
Different from semi-metallic
graphene\cite{NovoselovSci04,PKimNat05}, these systems display a
large band gap while still maintaining a high carrier
mobility\cite{Li2014,DT229,Koenig14,Xia14,Zant14}. Even though
phosphorus and arsenic are both group V elements, they crystallize
in different structures.
% The counterpart of white phosphorus, a molecular crystal
% consisting of P$_4$ molecules, is the poisonous yellow
% arsenic, a molecular crystal with As$_4$ constituents.
Most stable and thus more abundant are the layered allotropes such
as black phosphorus\cite{Bridgman14}, with the designation A17 or
$\alpha$-P, and grey arsenic\cite{DT244}, with the designation A7
or $\beta$-As.\cite{nomenclature}
%Zhen Zhu edit
% Ref. "nomenclature"
%The designation of $\alpha$- and $\beta$-phase to represents
%A17 and A7 is for the convenience of ordering layered black phosphorus
%allotropes with respect to their cohesive energy. This nomenclature is
%specially designed for layered group V elemental semiconductors and their
%isoelectronic counterparts, which should not be confused with other designations.
%end Zhen Zhu edit
Since a conversion of the $\alpha$ to the $\beta$
phase is possible\cite{DT230,Seifert2012}, combining both elements in the same
layer and changing the composition is bound to cause a structural
transition\cite{blackas,Krebs1957}. Since both structures are almost equally
stable, we may expect phase coexistence that should bring an
unexpected richness in both structural and electronic
properties\cite{{blackas},{blackastheory1},{blackastheory2},{BiluLiu15}}.
This way of isoelectronic doping could turn into an effective way
to fine tune the electronic properties,
%Zhen Zhu modify
improve the carrier mobility\cite{BiluLiu15}
%end Zhen Zhu modify
and reduce the chemical reactivity of the compound from those of
pristine of phosphorene and arsenene\cite{AsStable}.

% Executive summary

In this study we report \textit{ab initio} density functional
theory (DFT) calculations of As$_{1-x}$P$_x$ monolayers. We
determine the optimum geometry, the electronic structure, and
identify the most stable allotropes as a function of composition.
We predict a structural transition from the $\alpha$ to the
$\beta$ phase to occur near $x$=0.93. This structural transition
is accompanied by a change in frontier states from lone pair
electron states in $\alpha$-P to $\sigma$-bond states in
$\beta$-As and from a direct to an indirect fundamental gap.

% Here we present the relative stability and electronic properties
% of the As$_{1-x}$P$_x$ compounds. Based on our calculations, we
% find by varying the ratio of phosphorus component $x$, structural
% phase transition could be induced between $\alpha$-type (A17)
% structure (black phosphorus) and $\beta$-type (A7) structure (grey
% arsenic) near $x$=0.93. We also expect phase coexistence of both
% $\alpha$-type and $\beta$-type structures within larger range of
% $x$ due to their negligible stability difference. In $\alpha$
% phase, although the electronic band gap value almost does not
% vary, compounds with different mixing ratio of phosphorus and
% arsenic maintain either a direct fundamental band gap as in black
% phosphorus or an indirect band gap in $\alpha$-As). For the
% $\beta$ phase structures, the fundamental electronic band gap
% value depends significantly on the phosphorus component ratio $x$
% and the frontier states of will switch from lone pair electrons in
% $\beta$-P to the inlayer $\sigma$-bond electrons in grey arsenic.

%===========< FIGURE 1 >============================================
% entire page. There is no need to do explicit centering.
\begin{figure*}[t]
\includegraphics[width=1.6\columnwidth]{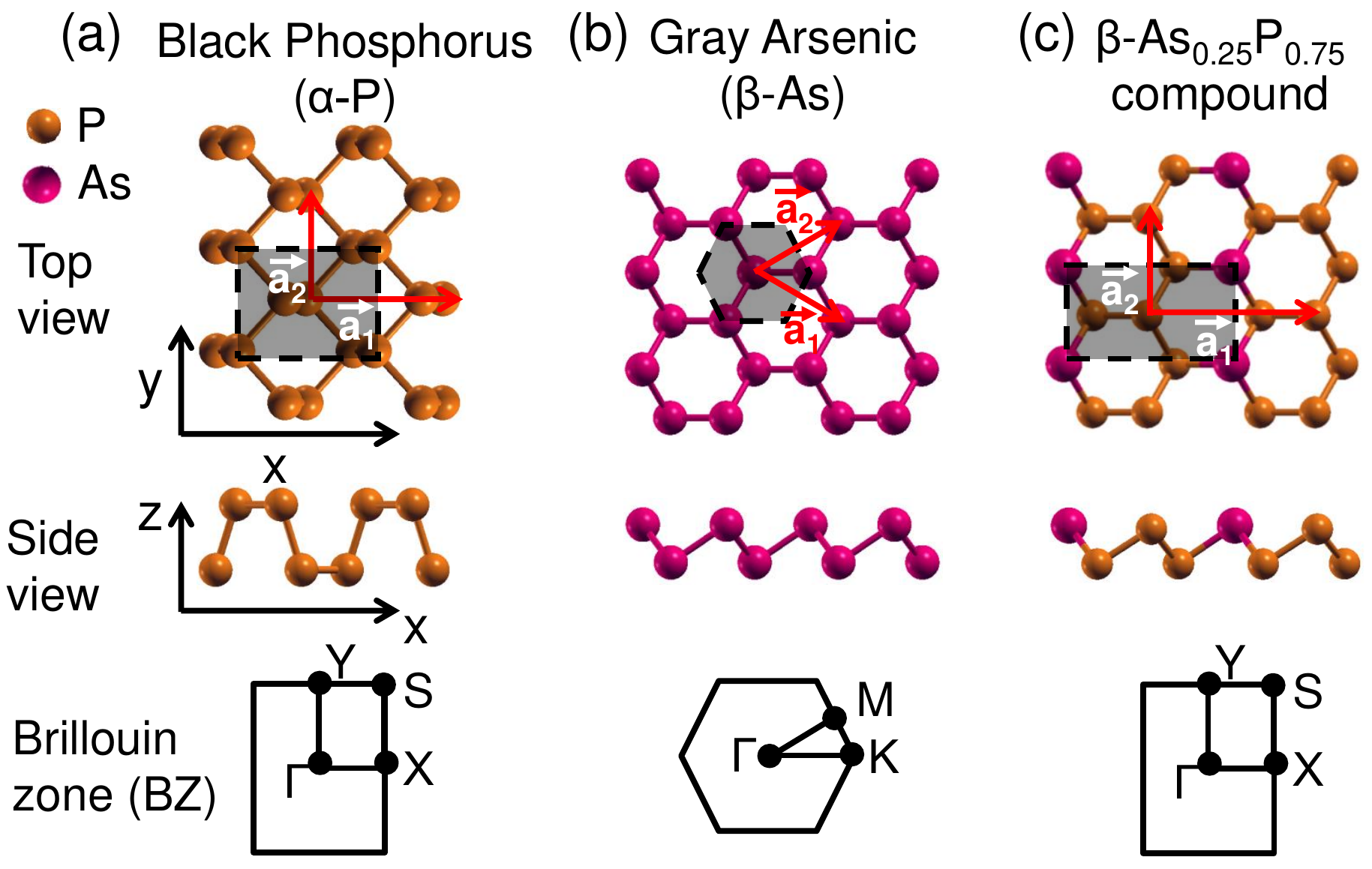}
\caption{(Color online) Equilibrium structure of a monolayer of
(a) black phosphorus ($\alpha$-P), (b) grey arsenic ($\beta$-As),
and (c) the $\beta$ allotrope of the As$_{0.25}$P$_{0.75}$
compound. Ball-and-stick models in top and side view, with the
primitive unit cell highlighted by shading, are shown in the upper
panels and the Brillouin zones in the lower panels. \label{fig1}}
\end{figure*}
%===========< FIGURE 1 >=========================================

%===========< FIGURE 2 >=========================================
% entire page. There is no need to do explicit centering.
\begin{figure*}[t]
\includegraphics[width=1.6\columnwidth]{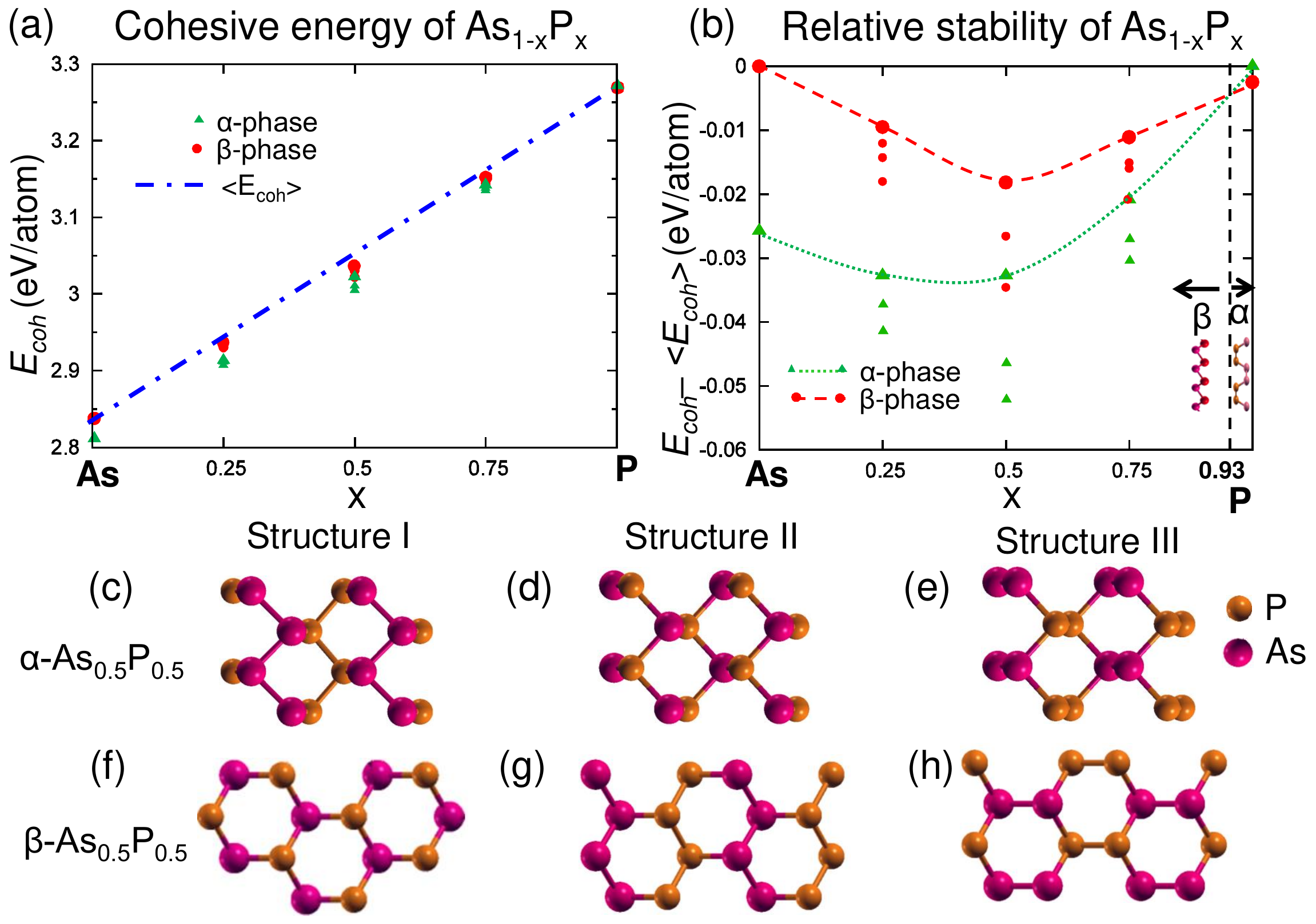}
\caption{(Color online) (a) Cohesive energy $E_{coh}$ of an
As$_{1-x}$P$_x$ monolayer as a function of the phosphorus
concentration $x$. The most stable allotropes are shown by larger
symbols. $<E_{coh}>$, shown by the dash-dotted line, is a linear
interpolation between the most stable allotropes of As and P. (b)
Relative stability $E_{coh}-<E_{coh}>$ of As$_{1-x}$P$_x$ as a
function of $x$. The vertical dashed line indicates the expected
composition for an $\alpha{\rightarrow}\beta$ structural
transition. Ball-and-stick models of selected stable structures of
As$_{0.5}$P$_{0.5}$ in the [(c) to (e)] $\alpha$ phase and [(f) to
(h)] $\beta$ phase. For each phase, the structural indices I, II
and III are arranged in the order of increasing stability.
\label{fig2}}
\end{figure*}
%===========< FIGURE 2 >======================================

\section{Results and discussion}

% structure

We present in the following computational results for the
equilibrium geometry and electronic structure of As$_{1-x}$P$_x$
compounds as a function of composition. The monolayer structures
have been optimized using DFT with the Perdew-Burke-Ernzerhof
(PBE)\cite{PBE} exchange-correlation functional, as discussed in
the Methods Section.
Since group V elements phosphorus and arsenic
are both threefold coordinated and display a tetrahedral bonding
character, they possess the freedom to arrange atoms in a layered
structure that is not flat, which leads to various buckled
allotropes that are topologically related to a honeycomb
lattice\cite{DT240}. Among these, the $\alpha$ and $\beta$ phases
are most stable\cite{DT230}. Under ambient pressure, phosphorus
favors the $\alpha$ and arsenic the $\beta$ structure, but an
$\alpha{\rightarrow}\beta$ transition has been reported under high
pressure\cite{Jamieson63,Seifert2012}. As seen in Fig.~\ref{fig1}(a), a
monolayer of $\alpha$-P (black phosphorus) can be viewed as a
distorted honeycomb lattice with a rectangular primitive unit
cell. The structure of a $\beta$-As (grey arsenic) monolayer,
presented in Fig.~\ref{fig1}(b), resembles more closely the
honeycomb lattice of graphene with a hexagonal unit cell.

We explored the structures of As$_{1-x}$P$_x$ compounds with
$0<x<1$ by considering different arrangements of As and P atoms in
systems with $x=0.25$, $0.5$ and $0.75$.
We optimized the lattice
structure for each system and found that the lattice constants
depend primarily on $x$ and increase with increasing As
concentration due to the larger atomic radius of arsenic in
comparison to phosphorus. In the strongly anisotropic $\alpha$
phase, we found an increase of 4\% along the softer $a_1$
direction and a 10\% increase along the harder $a_2$ direction
when moving from pristine phosphorus to arsenic. The corresponding
change in both $a_1$ and $a_2$ was an increase by 9\% in the
isotropic $\beta$ phase. The data for the dependence of the
lattice parameters on composition are presented in the Supporting
Information.

% stability and cohesive energy

Since the cohesive energy of phosphorus is larger than that of
arsenic, we expect a larger cohesive energy in phosphorus-rich
structures. In the first approximation, ignoring structural and
short-range order differences, we expect the cohesive energy of
the As$_{1-x}$P$_x$ compound to be a linear combination of the
cohesive energies of the pristine components in their respective
structures, $<E_{coh}>=(1-x)E_{coh}$(As)$+xE_{coh}$(P). In a
compound with a particular structure, we define the cohesive
energy per ``average atom'' by
$E_{coh}$(As$_{1-x}$P$_x$)$=-E_{tot}$(As$_{1-x}$P$_x$)$/N+
(1-x)E_{tot}$(As atom)$+xE_{tot}$(P atom), where
$E_{tot}$(As$_{1-x}$P$_x$) is the total energy of the $N$-atom
unit cell, $E_{tot}$(As atom) is the total energy of an isolated
As atom and $E_{tot}$(P atom) the total energy of a P atom. As
seen in Fig.~\ref{fig2}(a), the cohesive energy
$E_{coh}$(As$_{1-x}$P$_x$) will generally deviate from the
expectation value $<E_{coh}>$ for specific geometries and atomic
arrangements, but the deviations are rather small.

To better investigate these deviations, we define the relative
stability of a given structure by
$E_{coh}$(As$_{1-x}$P$_x$)$-<E_{coh}>$ and show the results in
Fig.~\ref{fig2}(b). In general, we find that combining both
elements in the same layer is always associated with an energy
penalty with respect to the $<E_{coh}>$ value, caused by the size
mismatch of As and P atoms. For a given composition, we found
cohesive energy differences between particular arrangements of As
and P atoms to be as large as 20~meV/atom at $x=0.5$. Since the
cohesive energy ranges of the different atomic arrangements within
the phases overlap, we expect a coexistence of the two phases in
realistic samples.

We found the monolayer of $\beta$-As to be more stable by
${\approx}26$~meV/atom than that of $\alpha$-As, whereas the
monolayer of $\alpha$-P is favored by ${\approx}3$~meV/atom over
$\beta$-P. As seen in Fig.~\ref{fig2}(b), the energy penalty in
the $\beta$ phase is smaller than that in the $\alpha$ phase for
lower $x$ values. Only in very phosphorus-rich compounds we find
the $\alpha$ phase to be more stable. Consequently, we expect a
structural transition between the $\alpha$ and the $\beta$ phase
to occur as a function of composition. Our results indicate that
this structural transition should occur near $x{\approx}0.93$.

To illustrate how the relative position of arsenic and phosphorus
atoms affects the stability of the compounds, we present three
different structures of As$_{0.5}$P$_{0.5}$ in the $\alpha$ phase
in Fig.~\ref{fig2}[(c) to (e)] and in the $\beta$ phase in
Fig.~\ref{fig2}[(f) to (h)]. As the cohesive energy of phosphorus
is larger than that of arsenic, P-P bonds are stronger than As-As
bonds. Consequently, the structure with the largest number of P-P
bonds should be most stable. On the other hand, maximizing the P-P
interaction would imply segregating P from As, which will strain
the structure. There is a trade-off between the two trends. For
the sake of convenience, we called the least stable structural
arrangement ``structure I'' and the most stable arrangement
``structure III'' in Fig.~\ref{fig2}. As seen in
Fig.~\ref{fig2}(e) and \ref{fig2}(h), the two most stable
structures in either phase contain an alternating arrangement of
isolated P-P and As-As dimers. In this arrangement, the strain
caused by different bond lengths can be minimized, while still
keeping at least some P-P bonds. Strain is also low in the
$\alpha$-II and $\beta$-I structures, but absence of P-P bonds
makes them less stable. The largest number of P-P bonds is
realized in the $\alpha$-I and $\beta$-II structures, but the
large strain energy caused by the coexistence of phosphorus and
arsenic chain arrangements makes these structures energetically
unfavorable.

%===========< FIGURE 3 >=========================================
% entire page. There is no need to do explicit centering.
\begin{figure*}[t]
\includegraphics[width=1.6\columnwidth]{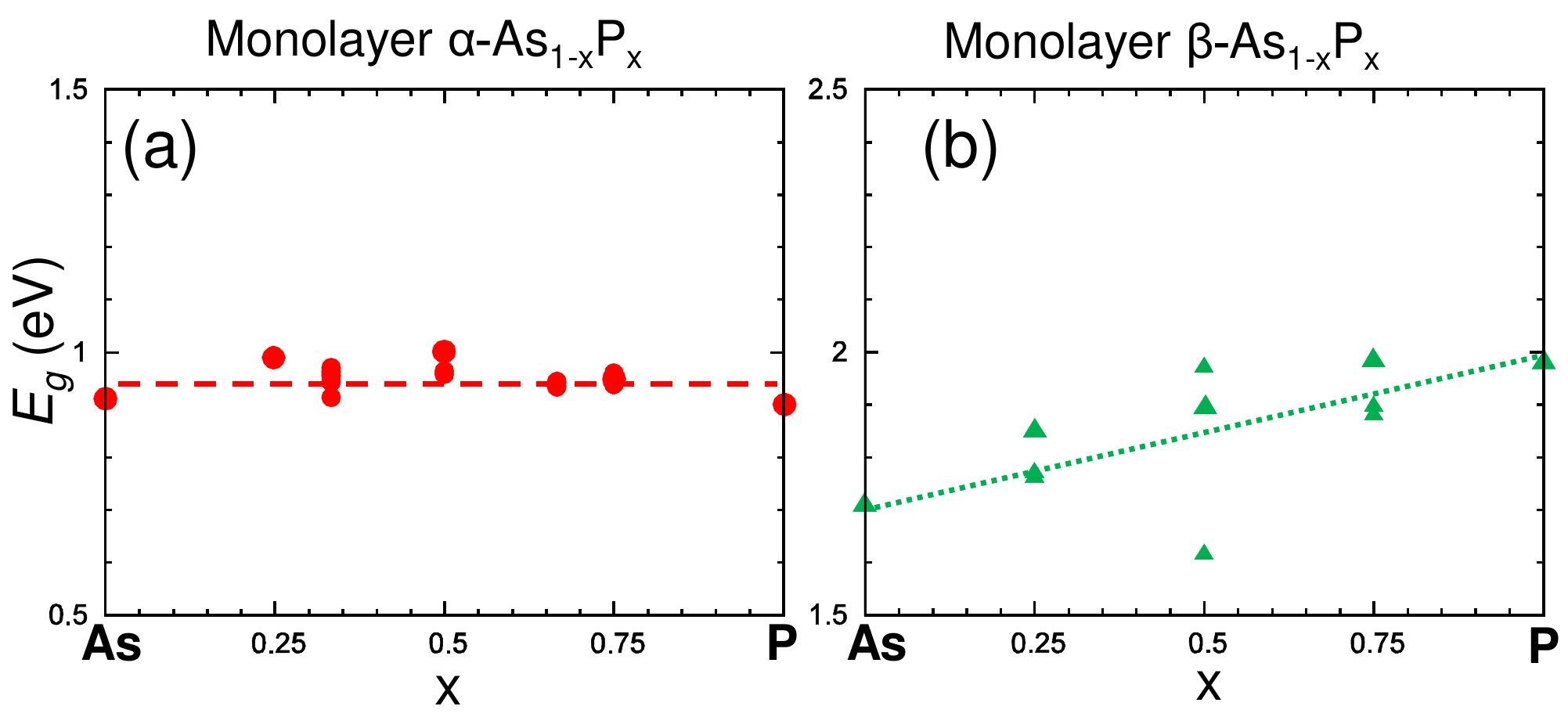}
\caption{(Color online) Fundamental electronic band gap $E_g$ in
As$_{1-x}$P$_x$ monolayers. DFT-PBE values of $E_g$ are presented
as a function of composition for compounds in the (a) $\alpha$-
and (b) $\beta$-phase. The dashed lines are guides to the eye.
\label{fig3}}
\end{figure*}
%===========< FIGURE 3 >=========================================

%===========< FIGURE 4 >=========================================
% entire page. There is no need to do explicit centering.
\begin{figure*}[t]
\includegraphics[width=1.8\columnwidth]{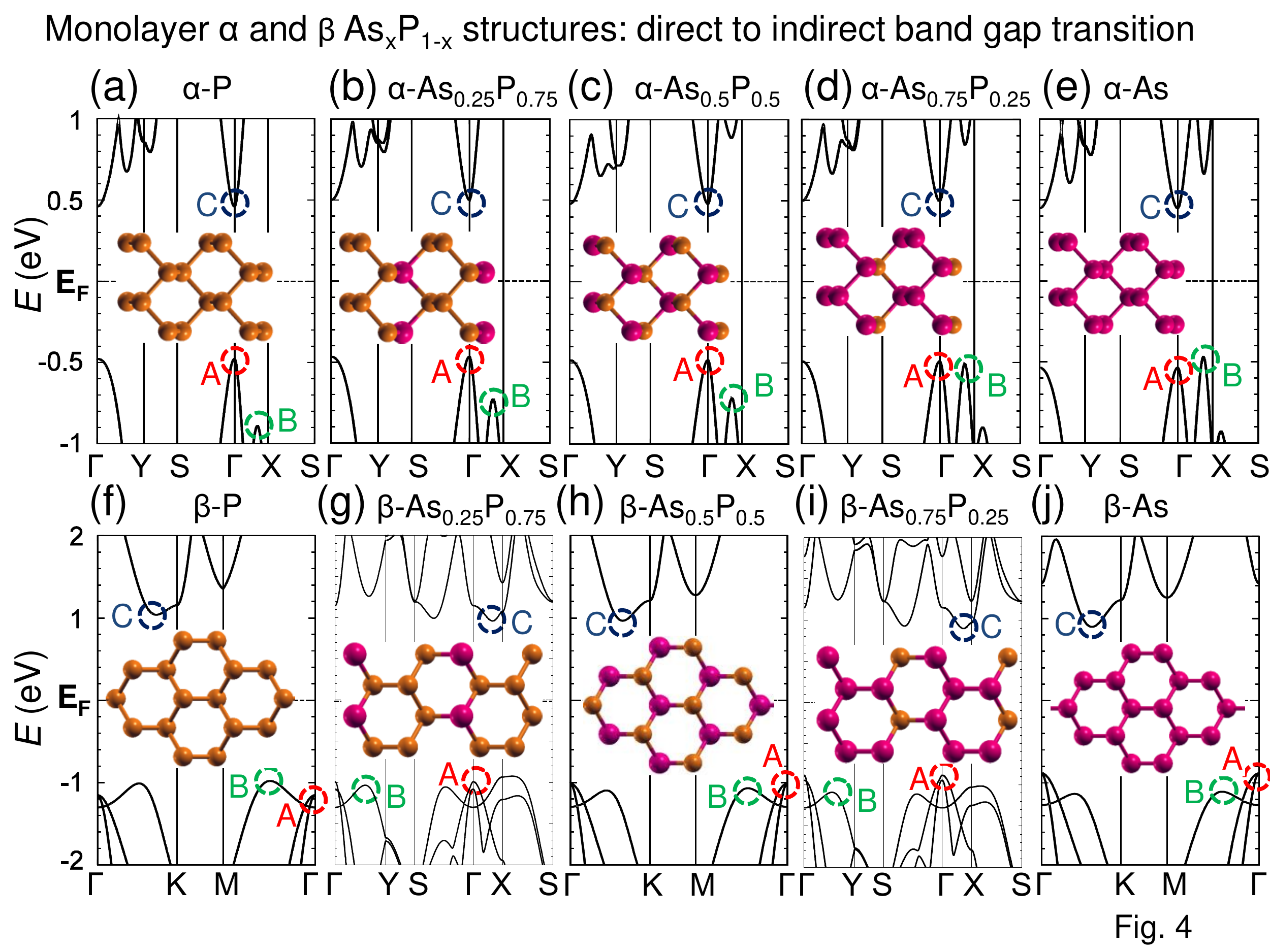}
\caption{(Color online) Electronic band structure of
As$_{1-x}$P$_x$ monolayers in the $\alpha$-phase [(a) to (e)] and
the $\beta$-phase [(f) to (j)]. The structural arrangements for
each composition are shown in the insets. The changing role of the
valence band maxima, shown by the dashed circles labeled ``A'' at
$\Gamma$ and ``B'' off-$\Gamma$, is discussed in the text. The
position of the conduction band minima is indicated by
the dashed circles labeled ``C''.%
\label{fig4}}
\end{figure*}
%===========< FIGURE 4 >=========================================

% electronic properties

As mentioned above, the relatively small stability difference
between the $\alpha$ and $\beta$ phase of As$_{1-x}$P$_x$
compounds suggests a likely coexistence of different allotropes.
Understanding the electronic properties of such a complex system
requires obtaining information about every such structural
arrangement. We have performed the corresponding calculations and
show the fundamental electronic band gap of As$_{1-x}$P$_x$
compounds as a function of composition in Fig.~\ref{fig3}. As
indicated in Fig.~\ref{fig3}(a), the electronic band gap value of
the $\alpha$ phase does not depend sensitively on the phosphorus
concentration $x$ and lies in the range between $0.9-1.0$~eV. For
the $\beta$ phase of As$_{1-x}$P$_x$ compounds, the band gap is
much larger and it's value increases with increasing phosphorus
concentration, as seen in Fig.~\ref{fig3}(b). The band gap values
not only display a larger value range from $1.6-2.0$~eV, but also
depend on the relative arrangement of P and As atoms within the
unit cell at a given composition.

More interesting than the absolute value of the band gap are
changes in the band structure of As$_{1-x}$P$_x$ compounds caused
by changing composition, which should be correctly captured by
DFT-PBE. Especially important appears to be the
position of the valence band maximum, since these compounds are
expected to behave as $p$-type semiconductors. The band structure
of $\alpha$-phase compounds is shown in Fig.~\ref{fig4}(a) for
pristine phosphorus, Fig.~\ref{fig4}(e) for pristine arsenic, and
in Figs.~\ref{fig4}(b) through \ref{fig4}(d) for intermediate
compositions. Careful comparison of these band structure results
indicates a transition from a direct gap in $\alpha$-P to an
indirect gap in $\alpha$-As. For the sake of discussion, we found
it useful to identify the valence band maximum at $\Gamma$ as
point ``A'', another local valence band maximum along the
${\Gamma}-X$ line as point ``B'', and the conduction band minimum
at $\Gamma$ as point ``C''. With increasing concentration of As,
the valence band maximum switches from ``A'' to ``B'', with the
transition occurring near As$_{0.75}$P$_{0.25}$ as seen in
Fig.~\ref{fig4}(d). Consequently, the character of the fundamental
band gap in As$_{1-x}$P$_x$ compounds in the $\alpha$ phase is
expected to change from direct for $x{\gtrsim}0.75$ to indirect
for $x{\lesssim}0.75$. This behavior is reminiscent of the
direct-to-indirect gap transition in a pure $\alpha$-phosphorene
monolayer that is induced by tensile in-layer strain and may be
rationalized by the fact that the larger atomic radius of As atoms
causes such strain. In reality, we find that the relative
positions of the ``A'' and ``B'' peaks depend not only on the
composition, but also the relative arrangement of phosphorus and
arsenic atoms. We compare the electronic band structure of
different structural arrangements with $x$=0.5 in the Supporting
Information.

The band structure changes in $\beta$-As$_{1-x}$P$_x$ compounds
with changing composition are presented in Fig.~\ref{fig4}(f) to
\ref{fig4}(j). Similar to the $\alpha$-phase structures discussed
above, we label the valence band maximum at $\Gamma$ as point
``A'', another local valence band maximum away from ${\Gamma}$ as
point ``B'', and the bottom of the conduction band as point ``C''.
Even though -- unlike the $\alpha$-phase compounds -- all
$\beta$-phase structures are indirect-gap semiconductors, the
position of the valence band maxima and conduction band minima
still changes with composition. There is a change in
$\beta$-As$_{1-x}$P$_x$ from the top of the valence band being at
``B'' in P-rich compounds with $x{\gtrsim}0.5$ to ``A'' at
$\Gamma$ in As-rich compounds with $x{\lesssim}0.5$. Similar
changes in the valence frontier states have been observed in a
pure As monolayer under strain\cite{DT244}. Also here, we find
that point ``A'' corresponds to in-plane $\sigma$-bond states,
whereas point ``B'' corresponds to lone pair electron
states,\cite{DT244} same as in a $\beta$-P monolayer. In other
words, changing the composition allows to change the character of
the frontier states in the valence band region and thus to
effectively tune the electronic properties of the system.

In conclusion, we have performed \textit{ab initio} density
functional theory (DFT) calculations of As$_{1-x}$P$_x$ monolayers
in search of ways to further improve the electronic properties of
group V layered semiconductors and reduce their chemical
reactivity. We have determined the optimum geometry, the
electronic structure, and have identified the most stable
allotropes as a function of composition. We have found the most
stable allotropes to be based on the $\alpha$ (A17) structure of
black phosphorus and the $\beta$ (A7) structure of grey arsenic.
Since the stability difference between these two phases is very
small, we expect coexistence of $\alpha$- and $\beta$-type
structures within a large composition range. Based on our results,
we expect a structural transition from the $\alpha$ to the $\beta$
phase to occur near $x$=0.93. This structural transition should be
accompanied by a change in frontier states from lone-pair electron
states in $\alpha$-P to $\sigma$-bond states in $\beta$-As and
from a direct to an indirect fundamental gap.

\section{Methods}

Our computational approach to gain insight into the equilibrium
structure, stability and electronic properties of various
phosphorene structures is based on {\em ab initio} density
functional theory as implemented in the \textsc{SIESTA}
code.\cite{SIESTA} We used periodic boundary conditions throughout
the study. We used the Perdew-Burke-Ernzerhof\cite{PBE}
exchange-correlation functional, norm-conserving Troullier-Martins
pseudopotentials\cite{Troullier91}, and a double-$\zeta$ basis
including polarization orbitals. The reciprocal space was sampled
by a fine grid\cite{Monkhorst-Pack76} of
$8{\times}8{\times}1$~$k$-points in the Brillouin zone of the
primitive unit cell. We used a mesh cutoff energy of $180$~Ry to
determine the self-consistent charge density, which provided us
with a precision in total energy of $\leq$ 2~meV/atom. All
geometries have been optimized by \textsc{SIESTA} using the
conjugate gradient method\cite{CGmethod}, until none of the
residual Hellmann-Feynman forces exceeded $10^{-2}$~eV/{\AA}.

\begin{acknowledgements}
We thank Bilu Liu for useful discussions. This study was supported by
the National Science Foundation Cooperative Agreement
\#EEC-0832785, titled ``NSEC: Center for High-rate
Nanomanufacturing''. Computational resources have been provided by
the Michigan State University High Performance Computing Center.
The authors declare no competing financial interest.
\end{acknowledgements}

%\bibliography{AsP15}

\begin{mcitethebibliography}{28}
\bibitem[nom()]{nomenclature}
The designation of $\alpha$- and $\beta$-phase to represents A17 and A7 is for
  the convenience of ordering layered black phosphorus allotropes with respect
  to their cohesive energy. This nomenclature is specially designed for layered
  group V elemental semiconductors and their isoelectronic counterparts, which
  should not be confused with other designations.\relax
\mciteBstWouldAddEndPunctfalse
\mciteSetBstMidEndSepPunct{\mcitedefaultmidpunct}
{}{\mcitedefaultseppunct}\relax
\EndOfBibitem
\end{mcitethebibliography}
%\end{document}
\providecommand{\latin}[1]{#1}
\providecommand*\mcitethebibliography{\thebibliography}
\csname @ifundefined\endcsname{endmcitethebibliography}
  {\let\endmcitethebibliography\endthebibliography}{}

%%%%%%%%%%%%%%%%%%%%%%%%%%%%%%%%%%%%%%%%%%%%%%%%%%%%%%%%%%%%%%%%%%%%%
\end{document}